# Antiferroelectric nanodomains in the vicinity of the morphotropic boundary in PZT-based coarse-grained ceramics

V. M. Ishchuk[1], D. V. Kuzenko[1], and V. L. Sobolev[2]

[1]*Science & Technology Center "Reactivelectron" of the National Academy of Sciences of Ukraine, Donetsk 83049, Ukraine*
[2]*Department of Physics, South Dakota School of Mines & Technology, Rapid City, South Dakota 57701*

Results presented in this article demonstrate importance of taking into account such a phenomenon as the solid solution decomposition at the boundaries separating coexisting phases in lead-zirconate-titanate-based solid solutions with compositions belonging to the morphotropic boundary region of the "temperature-composition" phase diagram. It is shown that in the local decomposition of solid solutions in the vicinity of the boundaries separating the tetragonal and rhombohedral phases in lead-zirconate-titanate-based solid solutions lead to the changes of the chemical composition of the solid solutions and to the formation of segregates. It is also shown that the proper thermoelectric treatment of samples containing these segregates can give substantially higher values of piezoelectric parameters in the lead-zirconate-titanate-based compounds.

PACS number(s): 77.90.+k , 77.84.Cg, 77.80.B-, 77.80.bg, 77.80.Dj, 81.30.Dz.

# 1. ITRODUCTION

An observation of the monoclinic phase in lead zirconate-titanate (PZT) solid solutions with compositions belonging to the so-called morphotropic boundary (MPB) region of the "composition-temperature" diagram of phase states was first reported in [1 - 3]. After appearance of these articles a large number of publications emerged either reporting confirmation of the presence of the monoclinic phase in PZT or trying to give theoretical explanation of observed results using phenomenological approach or first principle calculations. We simply cannot cite all these papers here since the list of published studies would exceed the size of any original publication. All the results of research done on this topic are given and analyzed in a series of reviews [4 - 7], see also collection of papers [8]. In these reviews, one can find all references on original studies devoted to investigations of the MPB region in PZT.

Let us mention fundamental peculiarities of publications on a monoclinic phase in PZT in our opinion. Results of crystal structure investigations have been based on the analysis of profiles of the lines obtained by means of X-ray or neutron scattering. The analysis itself has been based on the models that take into account concrete individual crystal structures (these are the tetragonal, the rhombohedral, and three monoclinic structures in the case of PZT) or their coexistence in the bulk of the sample under investigation. The structure for which the calculated profile of the lines has been the closest to the experimentally obtained profile has been considered as a "true" structure.

As it follows from the above-mentioned procedure, only the effects and phenomena that have been well known previously, as well as the types of the crystal structures can be obtained by using this way of data analysis in X-ray and neutron diffraction measurements. Any phenomenon with which the individuals analyzing data have not been familiar has not been taken into account and does not play any role in final solution of the problem.

In the case of PZT and PZT-based solid solutions such unaccounted phenomenon is the local decomposition of the solid solution in the vicinity of the coherent interphase boundaries (CIPB). The boundaries between the domains of the tetragonal and rhombohedral phases coexisting within one single-crystalline grain are exactly such CIPB boundaries. These two coexisting phases are the phases with different inter-plane distances and different parameters of their crystal lattices. In the immediate vicinity of such CIPBs the local decomposition of solid solution takes place to lower the elastic energy. The ions with larger ionic radii are expelled into the domain with larger parameters of the crystal lattice and the ions with smaller ionic radii are moved into the domain with smaller crystal lattice parameters (the ions in equivalent crystallographic positions are referred to here). Due to this decomposition, the chemical composition of solid solutions in the vicinity of these interphase boundaries differs from the chemical composition inside the domains of both phases. As a result the segregates with the crystal lattice close to the parent solid solution composition but still different are formed. Such local decomposition of the solid solution in the vicinity of coherent interphase boundaries in PZT-based solid solutions has been detected and studied in details in [9 - 12].

X-ray and neutron diffraction investigations of the crystal structure of PZT solid solutions from the MPB region of compositions represent the larger group of published studies. These studies are based on the analysis of the profiles of lines of the diffraction pattern due to the elastic scattering of incident radiation. The local decomposition of solid solutions and the formation of segregates are manifested in the appearance of additional lines in the X-ray and neutron diffraction patterns. These new lines are usually wider and weaker than the Bragg lines of the main perovskite crystal lattice and are located at the angles close to the corresponding angles of the Bragg lines. The volume occupied by the new phases is small. The crystallographic planes of these regions of new phases are randomly located, and as a result, the corresponding X-ray lines have appearance of diffuse lines. This means that the profiles of the X-ray or neutron diffraction lines appear somewhat changed which can be attributed to the appearance of new phases. As far as we know, the local decomposition of solid solution has never been taken into account during the analysis of the crystal structure in the MPB region.

Spectroscopy methods are the basis for another group of studies of crystal structure of PZT-based solid solutions. The Raman light scattering has been used for the most part. Profiles of the spectral lines in the infrared absorption spectra (and lines in the reflection spectra) are used to evaluate the possible structural peculiarities of PZT solid solutions. The situation in this approach is analogous to the situation that took place in the treatment of data of the X-ray and neutron experiments. New lines (with a weak intensity) that appear because of the local decomposition of solid solution in the vicinity of the interphase boundaries lead to the modification of the profiles of lines. It can lead to the ambiguity in identification of the crystal structure of solid solutions if the above phenomenon has not been taken into account.

All said above can be referred to results of the electron diffraction. The presence of segregates may lead to the additional splitting of main electron diffraction reflexes as well as to appearance of additional reflexes.

The ions of zirconium and titanium are shifted inside the domains with different parameters of the crystal lattice in the process of the local decomposition. As a result local zirconium enriched regions appear. The antiferroelectric (AFE) state can be realized in these local regions. It is difficult to elicit directly such-like segregates with the AFE state in the PZT solid solutions at present. It has been done for the PZT-based solid solutions with compositions from ferroelectric-antiferroelectric morphotropic region of the "composition-temperature" phase diagram in [9, 11, 12, 15, 16, 20] (where the sizes of segregates were of the order of several tens of nanometer). In addition to that, several publications appeared recently that pointed out the possible existence of the AFE ordering in the PZT solid solutions with compositions from the MPB region (the region between the tetragonal and rhombohedral phases) of the phase diagram. Results obtained in these publications we will give during the discussion of our results.

The presence of segregates leads to a series of specific effects characteristic to the boundaries between coexisting domains with the AFE and FE orderings [13 - 16].

The goal of this study has been the search and investigation of such effects in the PZT-based solid solutions with the compositions from the MPB region of the "composition-temperature" phase diagram. In no case, we try to assert the absence of the monoclinic phase in the PZT solid solutions. We only want to show that there are physical effects that have not been taken into account during the interpretation of experimental results and during the identification of the crystal structure of solid solutions.

We have developed a procedure for manufacturing of the PZT ceramics (with different ion substitutions) with compositions located near the boundary between the tetragonal and rhombohedral phases in the "composition-temperature" phase diagram, with a grain size exceeding 20 μm. Manufactured ceramic samples are characterized by a high degree of the composition homogeneity in the bulk of the substance. This has been confirmed, first of all, by the grain size and, secondly, by the extremely small width of the MPB region (less than 0.5%) in the investigated seven-component PZT-based solid solutions with compositions $(Pb_{0.9}Ba_{0.05}Sr_{0.05})_{0.985}(Li^{+}_{0.5}La^{3+}_{0.5})_{0.015}(Zr_{1-x}Ti_x)O_3$.

According to the data available in literature, the grain size of PZT ceramic samples usually used in experiments is of the order of several micrometers. That is why the rhombohedral and tetragonal phases are in different grains and up to a recent time it has been very difficult (or even completely impossible) to investigate effects caused by the interphase domain boundaries. In this case, grain boundaries play the role of the interphase boundaries. It is common knowledge that inter-grain boundaries accumulate a lot of impurities as well as lattice imperfections due to impurities. It is understood that there is no gradual conjugation of crystal planes in this case. The sizes of crystal grains in our samples have been of the order 30 to 40 microns. The domains of two phases can coexist within one crystallite separated by the interphase boundaries (CIPBs) of the atomic scale and character. There were no discontinuities of crystal plains (we have seen it using TEM) and the boundaries had coherent character without dislocations and the concentration of the elastic stress. A high degree of samples' homogeneity has allowed us to study the kinetics of development of segregates in the vicinity of CIPBs [10, 11, 16] along with the physical effects caused by these CIPBs.

In our opinion it is important to note one more peculiarity of the absolute majority of publications on monoclinic phase in PZT-based solid solutions with compositions from the MPB region of the phase diagram. Because of the wide region of coexistence of phases, the exact phase boundaries between various phases cannot be located using diffraction data. The wide region of phase coexistence is a convincing characteristic feature of the inhomogeneity of samples. The last circumstance is an additional factor contributing to the uncertainty in determination of the crystal structure of solid solutions.

## II. EXPERIMENTAL METHODS

The $(Pb_{0.9}Ba_{0.05}Sr_{0.05})_{0.985}(Li^{+}_{0.5}La^{3+}_{0.5})_{0.015}(Zr_{1-x}Ti_x)O_3$ solid solutions with the (Zr/Ti) compositions located in the regions of both the rhombohedral and the tetragonal distortions of the perovskite crystal lattice as well as with compositions belonging to the MPB region of the "composition-temperature" phase diagram have been used in our studies.

Disk-shaped ceramics samples of standard size $d = 10$ mm, $h = 1$ mm were manufactured using modified ceramic technology via the two-stage sintering at 850 °C and 1200 °C. The residual porosity of samples was not more than 0.2.

A control of the single phase structure of samples and the structural analysis was performed using SIEMENS D-500 powder X-ray diffractometer with *Ge* monochromator ($CuK_{\alpha_1}$-radiation with a wavelength of 1.54056 Å) on primary beam and BRAUN gas position-sensitive detector. The measurement conditions for obtaining X-ray lines profile provided 0.01° resolution with respect to the angle 2θ.

Studies of sample's microstructure were carried out using a JSM-820 scanning electron microscope with a Link-10000 attachment for control of chemical composition of samples. The electron microscopy data showed that the grain sizes of samples were of the order of 30 − 40 μm. Silver electrodes were used for dielectric and piezoelectric measurements.

The thermal treatment of sample was conducted in the low-inertia electric furnace with the chromel-alumel thermocouple. The heating temperature was varied from 20 to 500°C. Isothermal endurance (when it was necessary) was carried out with precision ± 0.75°C.

Temperature dependencies of the dielectric constant, ε(*T*), were measured in AC electric field with frequency of 1 kHz using precision QuadTech 7600 LCR meter. The poling of ceramic samples was carried out at a temperature of 120 °C in DC electric field with an intensity of 3 kV/mm during 1 h with the subsequent cooling down to room temperature in the electric field. Measurements of the piezoelectric module $d_{33}$ were performed by the quasi-static method [17].

## III. EXPERIMENTAL RESULTS

### A. Peculiarities of properties of coarse-grained ceramics

The "temperature-composition" phase diagram for the solid solutions with compositions $[(Pb_{0.9}Ba_{0.05}Sr_{0.05})]_{0.985}(Li^+{}_{0.5}La^{3+}{}_{0.5})_{0.015}(Zr_{1-x}Ti_x)O_3$ is shown in Fig.1 [18].

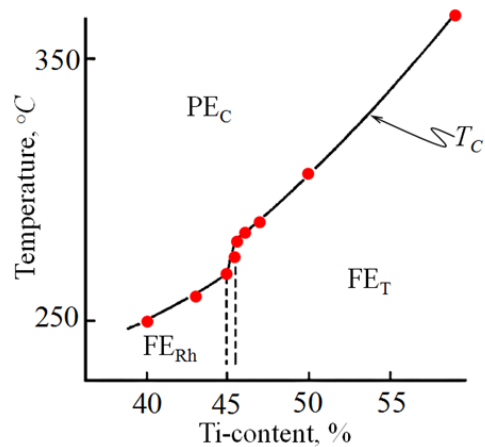

Fig.1. "Temperature-composition" phase diagram for $[(Pb_{0.9}Ba_{0.05}Sr_{0.05})]_{0.985}(Li^+_{0.5}La^{3+}_{0.5})_{0.015}(Zr_{1-x}Ti_x)O_3$ solid solutions in the vicinity of the MPB region of compositions.

Profiles of the (200) and (222) X-ray lines for the above solid solutions with compositions from the MPB region of the phase diagram are presented in Fig. 2. The coexisting domains with different types of crystal lattice distortions are present in the bulk of these samples. Availability of such coarse-grained ceramics (let us recall, that the size of ceramic grains have been in the range of 30 – 40 μm) made it possible to investigate specific effects caused by the coexistence of domains with the FE and AFE ordering within a single crystalline grains.

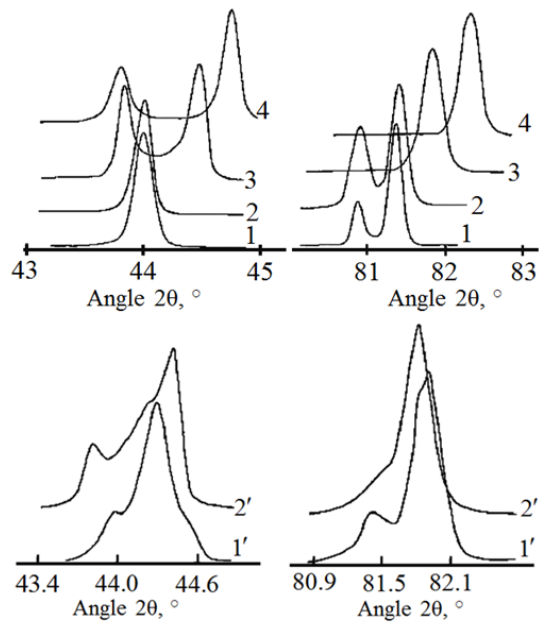

Fig.2. (200) and (222) X-ray diffraction lines for $[(Pb_{0.9}Ba_{0.05}Sr_{0.05})]_{0.985}(Li^+_{0.5}La^{3+}_{0.5})_{0.015}(Zr_{1-x}Ti_x)O_3$ solid solutions with different content of titanium.
Titanium content, x, %; 1 – 0.35; 2 – 0.45; 3 – 0.46; 4 – 0.47; $1'$ – 0.4550; $2'$ – 0.4575.

The concentration dependencies of the main dielectric and piezoelectric parameters of the $(Pb_{0.9}Ba_{0.05}Sr_{0.05})_{0.985}(Li_{0.5}La_{0.5})_{0.015}(Zr_{1-x}Ti_x)O_3$ solid solutions are given in Fig.3. As it seen from these dependencies, all parameters have a pronounced maximum located in the MPB region of Zr/Ti compositions. On the one hand, this fact confirms the made earlier conclusions about the nature of behavior of these parameters in the vicinity of MPB. However, a special attention has to be paid to the following fact. The MPB region of solid solutions under consideration is very narrow (not exceeding 0.5%), but the interval within which the parameters change is wide. In particular, for the coefficient of the electromechanical coupling, $K_r$, dielectric permittivity, $\varepsilon$, and polarization this interval is of the order of 10%. At the same time, the interval of high values of the piezoelectric modulus $d_{33}$ is narrow.

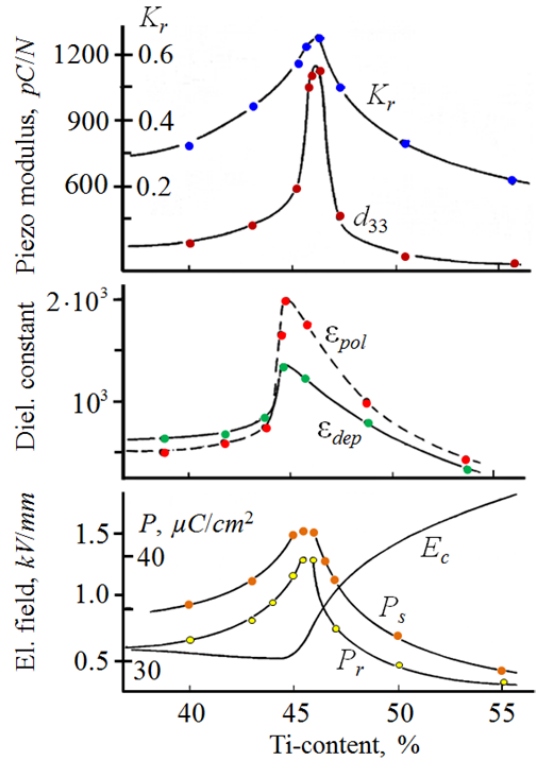

Fig.3. Composition dependencies of main physical parameters for $(Pb_{0.9}Ba_{0.05}Sr_{0.05})_{0.985}(Li_{0.5}La_{0.5})_{0.015}(Zr_{1-x}Ti_x)O_3$ solid solutions.

It should be noted that the values of the piezoelectric modulus $d_{33}$ are extraordinarily high for our PZT-based ceramics (more than $1100 \cdot 10^{-12}$ pC/N) with compositions belonging to the MPB region of the phase diagram. For comparison we have to remind that for well-known piezoelectric ceramic materials the value of piezoelectric coefficient $d_{33}$ does not exceed $600 \cdot 10^{-12}$ pC/N.

As a rule, extreme values of the parameters of PZT solid solutions are explained in the literature by the coexistence of the tetragonal and rhombohedral FE phases. It is generally considered that the polarization, dielectric, and piezoelectric parameters of solid solutions from the MPB region are defined by the so-called orientation polarization.

However, we have obtained PZT-based solid solutions for which the width of the MPB region is less than 0.5%, but the interval of increase of the main physical parameters exceeds 10%. It is clear that in the present case the above-mentioned mechanism cannot explain such behavior. Taking into account the new (monoclinic) phase discussed recently in the literature does not help either, since the concentration interval of its existence is narrow (between the regions of tetragonal and rhombohedral phases). Such situation requires searching for new approaches to explanation of properties of those solid solutions which are located within the limits of the MPB region.

## B. Diffuseness of the phase transition from paraelectric to ordered state.

In what follows we will consider effects that take place in the substances with a small difference in energies of the FE and AFE states and as a result the coexisting domains of these phases are present. We will compare these effects with those that have been observed in the coarse-grained PZT ceramics with compositions belonging to the MPB region of the phase diagram.

In the case of the coexisting domains of the FE and AFE phases in the sample one has to take into account the interaction between these domains. In the simplest case of the equal stability of these phases (here the equality of the free energies of these phases is meant) in the frames of phenomenological Landau-Ginsburg theory the energy describing this interaction can be represented in the following form [13, 14, 16]:

$$W_{int} = \xi_1\xi_2 C_1 P_1 P_2 + \xi_1\xi_2 D_2 \eta_1 \eta_2 \quad (1)$$

The appearance of terms in this equation is discussed in [13]. The order parameters $P_1$ and $\eta_1$ correspond to the domain of the FE phase while $P_2$ and $\eta_2$ are the order parameters of the AFE phase, $\xi_1$ and $\xi_2$ are the shares of the FE and AFE phases in the crystal volume. Under the condition of equal stability of these phases $\xi_1 = \xi_2 = 1/2$. The coefficients $C_1$ and $D_2$ depend on the distance from the interphase boundary. Both $C_1$ and $D_2$ decrease when the distance from the boundary toward the interior of the domain of any of the phases increases. Due to this the interaction is essential in the vicinity of the boundaries and its influence decreases when the distance between the boundary and the domain interior increases. This interaction gives the energy gain for the zone in the vicinity of the boundary (this zone is determined by the region where coefficients $C_1$ and $D_2$ have nonzero values) [13, 16]:

$$\Delta W = -\frac{1}{32}\left(\frac{D_2^2}{V_1}\eta_{2,0}^2 + \frac{C_1^2}{V_2}P_{1,0}^2\right), \quad (2)$$

Here $P_{1,0}$ and $\eta_{2,0}$ are the values of the order parameters inside the FE and AFE domains at the temperatures below the Curie temperature correspondingly. Coefficients $V_1$ and $V_2$ are positive and are determined by the slope of the four-minima thermodynamic potential in the vicinity of equilibrium values of order parameters $P_{1,0}$ and $\eta_{2,0}$ in the $P$-$\eta$ plane. Lowering of energy leads to renormalization of the coefficients in the Landau expansion from

$$\alpha_1 = \alpha_0\left(T - T_{c,f}\right); \qquad \beta_1 = \beta_0\left(T - T_{c,af}\right) \quad (3)$$

to:

$$\alpha_1 = \alpha_0\left[T - \left(T_{c,f} + \frac{1}{32}\frac{C_1^2(x,y,z)}{V_2}\right)\right]; \qquad \beta_1 = \beta_0\left[T - \left(T_{c,af} + \frac{1}{32}\frac{D_2^2(x,y,z)}{V_1}\right)\right] \quad (4)$$

In these formulas $\alpha_1$ and $\beta_1$ are the coefficients of the $P^2$ and $\eta^2$ terms in the Landau expansion for nonequilibrium thermodynamic potential. Let us remind that we discuss the case of equal stability of the FE and AFE phases and due to this $T_{c,f} \approx T_{c,af}$. As one can see that the interphase interaction leads to the increase of the Curie temperature in the vicinity of the inter-domain boundary. This increase can be substantial when the constants $V_1$ and $V_2$ are small. Such situation takes place in the case of gently sloped minima of thermodynamic potential. The diffuse character of the phase transition from paraelectric to ordered phase is the consequence of the dependence of the Curie temperature on coordinates in these local regions of the crystal. Such diffuse character of the phase transition depends on the position of the solid solution (the solid solution composition) in the "temperature-composition" diagram of phase states. If the solid solution composition is located far from the boundary separating regions of the FE and AFE states the domains of the less stable (from the energy point of view) phase are practically absent in the sample's volume so the regions with elevated values of the Curie temperature are also absent and, thus, the smearing of the phase transition is absent.

The degree of diffusion of the paraelectric phase transition for the substances, in which domains of the FE and AFE phases coexist, depends on the location of this substance on the diagram of phase states. The results of investigation of phase transitions for a number of PZT-based solid solutions, in which the presence of coexisting phases has been confirmed by the methods of the transmission electron microscopy and X-ray analysis of the crystal structure, are presented in [9, 19-24]. For the following comparison some of these results are shown in Figs.4, 5, 6. The parameter of the phase transition diffuseness, $\delta$, is defined by the formula

$$\frac{1}{\varepsilon} = \frac{1}{\varepsilon_{max}} + \frac{1}{2\varepsilon_{max}\delta^2}(T-T_{max})^\gamma$$

(here $T_{max}$ and $\varepsilon_{max}$ correspond to the maximum of $\varepsilon(T)$ dependence). The dependencies of this parameter on solid solution composition along with the "composition-temperature" and "pressure-temperature" phase diagrams are presented in these figures. As one can see the diffuseness of the phase transition is defined by the location of the composition of considered solid solution in the phase diagram. A considerable increase of the phase transition diffuseness is typical for those solid solutions, which are located in the vicinity of the boundary between the FE and AFE states in the corresponding phase diagram. Such behavior of the phase transition from the paraelectric phase into ordered state is characteristic of the substances, in which the domains of the FE and AFE phases coexist under certain conditions (namely, in the vicinity of the boundary separating regions of the FE and AFE states in the phase diagram).

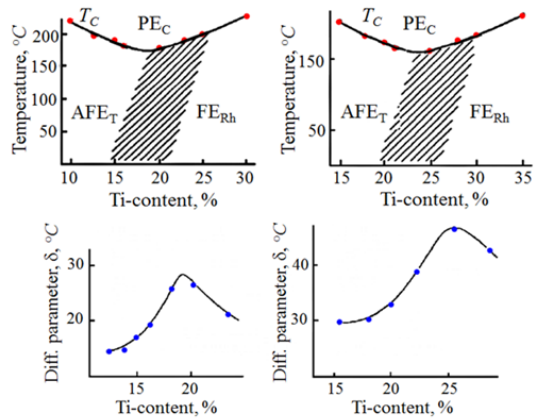

Fig.4. Phase diagrams of $Pb_{1-y}(Li_{0.5}La_{0.5})_y(Zr_{1-x}Ti_x)O_3$ solid solutions and corresponding dependencies of the diffuseness parameter in titanium content.
The content of $(Li_{0.5}La_{0.5})$ complex, %: On the left -10%, on the right-15%.

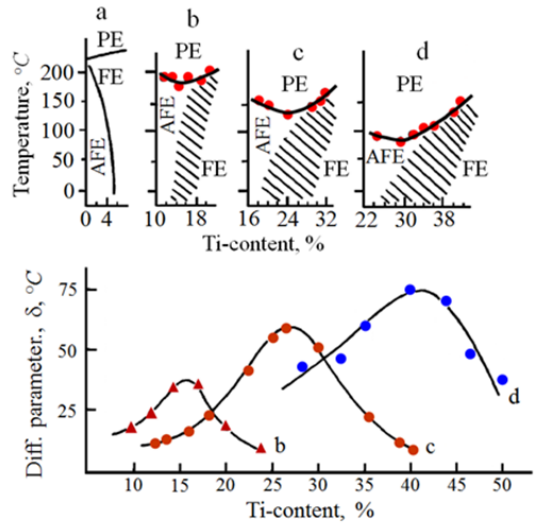

Fig.5. Phase diagrams of $Pb_{1-3y/2}La_y(Zr_{1-x}Ti_x)O_3$ solid solutions (upper part) and the dependencies of the diffusion parameter on the content of titanium (lower part).
The content of lanthanum, %: a – 0; b – 4; c – 6; d – 8.

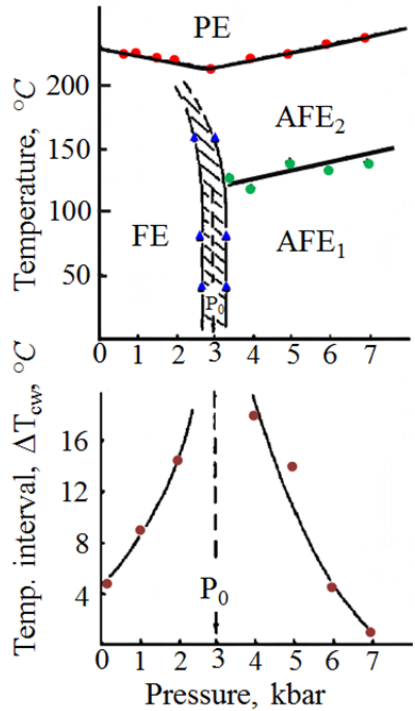

Fig.6. "Pressure-temperature" phase diagram of the PZT solid solution with 98/2 Zr/Ti-content (above) and the dependence of the temperature interval of diffuse paraelectric phase transition on pressure (below).

The dependence of the temperature interval, $\Delta T_{cw}$ within which the deviation from the Curie-Weiss law takes place on the Ti-content in $(Pb_{0.9}Ba_{0.05}Sr_{0.05})_{0.985}(Li_{0.5}La_{0.5})_{0.015}(Zr_{1-x}Ti_x)O_3$ solid solutions is given in Fig.7. This temperature interval characterizes the diffuseness of the paraelectric phase transition. As one can see the diffuseness of the paraelectric phase transition in this system of solid solutions also depends on the solid solution composition. The maximum diffuseness of the phase transition is observed in the solid solutions with Zr/Ti compositions belonging (according to the X-ray data) to the MPB region of the diagram of phase states.

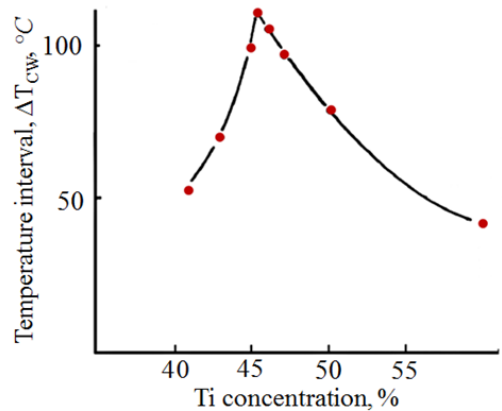

Fig.7. Temperature interval of diffuseness of the paraelectric phase transition vs. Ti-content in $(Pb_{0.9}Ba_{0.05}Sr_{0.05})_{0.985}(Li_{0.5}La_{0.5})_{0.015}(Zr_{1-x}Ti_x)O_3$ solid solutions

Such behavior of the diffuseness parameter of the paraelectric phase transition suggests a possible presence of domains with the AFE ordering in solid solutions with compositions falling within the MPB region.

### C. Local decomposition of solid solutions in the paraelectric phase

Consider now another effect manifested on coherent inter-phase boundaries between the domains of the FE and AFE states in the substances with a small energy differences between the these states.

The coexisting domains of the tetragonal and rhombohedral FE phases have different inter-planar distances. The transmission electron microscopy was used to demonstrate the coherent character of the CIPBs in PZT-based solid solutions in our earlier papers [22-24]. No concentration of deformations was revealed in the vicinity of CIPBs. This allowed us to make a conclusion about the inner structure of the said boundaries. The transition through CIPB (the given CIPB may be called seed or 'bare') is accompanied by continuous conjugation of the crystal planes. Such a coherent behavior must be accompanied by an increase of the elastic energy of the lattice along these boundaries.

In the said substances equivalent sites of the crystal lattice are occupied by ions, which have different sizes and/or different charges. Within a single-phase domain (far from the boundaries) each ion is not subjected to the influence of forces (in other words, the resulting force affecting each ion is equal to zero). Another situation takes place in the vicinity of CIPB. The balance of forces is disturbed. The 'larger' ions are pushed out into the domains with higher configuration volume, and consequently, with larger inter-planar distances. The 'smaller' ions are pushed out into the domains with lesser inter-planar distances. Such a process is accompanied, on one hand, by a decrease in the elastic energy along the CIPB, and, on the other hand, by an increase in the energy caused by the deviation of the solid solution composition from the equilibrium one. This process will end as soon as the structure of the new CIPB corresponds to the minimum of energy. This CIPB is now 'clothed'. The term CIPB will be used further just for such type of boundaries. In the considered solid solutions the A-sites of perovskite crystal lattice are occupied by ions with different ionic charge ($Pb^{2+}$, $Ba^{2+}$, $Sr^{2+}$, $La^{3+}$, $Li^{+}$), therefore, the local decomposition of the solid solution along CIPB may be accompanied by the local disturbance of electronegativity. Thus, the formation of the heterophase structure is accompanied by the disturbance of samples' chemical homogeneity i.e. by the local decomposition of solid solutions in the vicinity of CIPBs. At high temperatures, when the phases do not coexist, the samples remain homogeneous.

In what follows we present experimental verification of the above-described model for PZT-based solid solutions with compositions within the MPB region. We have identified effects that can be revealed experimentally and then the appropriate experiments have been carried out.

The local decomposition of the PZT-based solid solution in the vicinity of CIPBs leads to the appearance of two types of local domains. One type is the local domain enriched with lead zirconate and the other type is the local domain enriched with lead titanate. The latter type of titanium enriched domains as well as the whole matrix must exist in the FE state at the temperature below the Curie point. This is the reason why it is difficult to reveal their manifestation against the general background. However, the domains enriched with lead zirconate may be in the AFE state which cardinally differs from the FE state. Thus within the MPB region of solid solution compositions there exist specific FE+AFE domains formed as a pair of these phases as a result of the local decomposition of the solid solution. What is more the FE part of these domains has higher content of titanium than the matrix. This system of domains of coexisting phases manifests a number of specific effects, which are not characteristic of other dipole ordered systems which give an opportunity to observe them experimentally.

It follows form equations (2), (3), and (4) that the complex (FE+AFE) domain with inhomogeneous structure can exist at the temperatures above the temperature of the paraelectric phase transition inside the single-phase domains. These complex domains appear as a pair

FE+AFE phases separated by the interphase boundary. Appearance of these domains happens under definite conditions and in the solid solutions with particular compositions which we have learned to proportion. The temperature interval above the Curie point where these domains may appear can reach and even exceed 100 °C. The discussed existence of complex domains above the Curie temperature has been first revealed experimentally by X-ray method in [20] (see also [15]). This implies that composite domains with the FE-AFE interphase boundary arise in the paraelectric state of the substance's matrix at cooing down from high temperatures. As it was described above the local decomposition of the solid solution takes place in the vicinity of these boundaries and the segregates appear (in a general case, these segregates are ferroelectrically active).

One can easily control the microstructure of solid solutions since the inhomogeneous structure of the FE-AFE domains and segregates in the vicinity of the interphase boundaries appear in the process of the phase transition which itself can be easily controlled by the electric field or pressure. Considered physical process allows creation of textured domain structures at already in the paraelectric phase with the help of an external electric field applied to a sample at the temperatures far above the Curie point.

In the case when the solid solution composition is located in the section of the phase diagram far from the MPB region (and the decomposition of this solid solution does not take place) it can remain in macroscopically depolarized at the room temperature in the aftermath of special thermoelectric treatments carried out at the temperatures well above the point of transition into dipole-ordered state. Such treatments can be the cooling of samples in the presence of an electric field or the isothermal annealing in the presence of a field. Piezoelectric coefficients of the solid solutions subjected to these treatments are equal to zero. Such behavior is characteristic for all ordinary ferroelectrics.

A different picture of behavior is characteristic for substances with roughly equal stability of the FE and AFE states. The two-phase domains that appear in the paraelectric matrix of samples during cooling down from high temperatures have a random distribution of spontaneous deformations and, correspondingly, random orientations of axes of spontaneous polarization in the absence of external electric field. Interphase boundaries also have a random distribution of their planes. The process of diffusional formation of segregates along the interphase boundaries fixes such random distribution of the places of these boundaries during further cooling. The growth of volume occupied by the two-phase domains also takes place during the lowering of temperature. In the vicinity of $T_C$ the volume of the whole sample undergoes the transformation into the inhomogeneous state of the coexisting FE and AFE phases with the directions of spontaneous polarization imposed by these segregates. The degree of polarization that appears under action of external electric field on such a sample will be small and the level of the piezoelectric properties will also be insignificant.

But if a cooling from the high temperatures is carried out in an electric field the picture is totally different. The distribution of the axes of spontaneous polarization (consequently, the distribution of direction of the interphase FE-AFE boundaries) of the coexisting two-phase domains will be preassigned by the direction of the field. In this case the cooling of samples leads to the formation of a texture, which is stabilized by the segregates at the interphase boundaries at room temperature (and at the temperatures close to it). The samples are microscopically polarized; therefore, it will be easy to induce piezoelectric resonance in these samples. The piezoelectric moduli of these samples are non-zero even if an additional polarization of the samples is not realized. When additional polarization (along the direction of the field) is produced by the external field in the process of thermoelectric treatment, the materials will possess a set of properties that cannot be achieved by means of any other kinds of treatment.

The behavior of piezoelectric activity is even more interesting when the samples have underwent the thermoelectric treatment at the temperatures above the Curie point and these samples have been cooled in the absence of the field.

The distribution of axes of spontaneous polarization of the complex (FE+AFE) domains and the distribution of directions of the interphase FE-AFE boundaries and segregates have been dictated by the direction of the field during this thermoelectric treatment. After electric field has been switched-off, the already oriented system of segregates preserves the ordered structure of the (FE+AFE)-domains in the paraelectric matrix of the substance (at $T > T_c$). After cooling the sample to temperatures $T < T_c$ the ordered domain structure stays preserved, and the sample manifests the piezoelectric activity (thanks to segregates only).

An example of described behavior is demonstrated in Fig.7 in the case of PZT-based solid solutions with the $Pb_{0.90}(Li_{0.5}La_{0.5})_{0.010}(Zr_{1-x}Ti_x)O_3$ composition [15]. The interval of titanium concentrations within which domains of the FE and AFE phases coexist in the bulk of the sample is located near x ≈ 19 – 20 (the phase diagram of this series of solid solutions is presented in Fig.4). The maximum diffuseness of the phase transition is manifested at this Ti-content.

Two types of thermoelectric treatment of samples were used in our studies. The first type of thermal treatment was carried out as follows. The sample was cooled from 500 °C to a temperature of 300 °C. DC electric field was applied to the sample at 300 °C and it was cooled in the presence of the field. When the sample was cooled down to the temperature $T_C(x) + 20$ °C the electric field was switched-off and after cooling down to room temperature the piezoelectric coefficient $d_{33}$ was measured. The $d_{33}(x)$ dependence for the first type of the thermoelectric treatment is shown in Fig. 8 (curve 2). During the second type of the thermoelectric treatment the sample was cooled to the temperature $T_C(x) + 20$ °C without electric field and a DC electric field was applied to the sample at this temperature and the sample was held under the action of this field for 1 hour. After one hour the electric field was switched-off, the sample was cooled down to room temperature and then the piezoelectric coefficient $d_{33}$ was measured. Dependence $d_{33}(x)$ for this type of the thermoelectric treatment is shown in Fig. 8 (curve 3). As can be seen after samples were subjected to the action of the field in the paraelectric state the piezoelectric activity is manifested only in solid solutions with compositions that possess the system of coexisting domains of the FE and AFE phases.

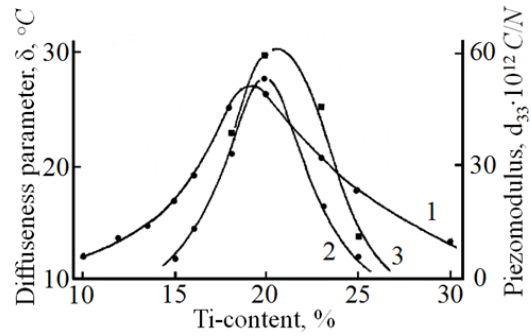

Fig.8. Dependence of the diffuseness parameter (1) and the piezoelectric moduli (2 and 3) measured after different regimes of the thermoelectric treatment in PE phase on the content of the $Pb_{0.90}(Li_{0.5}La_{0.5})_{0.010}(Zr_{1-x}Ti_x)O_3$ solid solution.

Let us now discuss PZT-solid solutions with compositions belonging to the MPB region of the phase diagram.

The results obtained in analogous experiments on solid solutions with $(Pb_{0.9}Ba_{0.05}Sr_{0.05})_{0.985}(Li_{0.5}La_{0.5})_{0.015}(Zr_{1-x}Ti_x)O_3$ compositions are presented in Fig.9. In this case the thermoelectric treatments have been performed at the temperatures above the Curie point and an electric field has been applied to samples at temperatures $T_C(x) + 50\ ^{\circ}C$ and the samples have been kept in the field during 20 min (curve 1) and 40 min (curve 2). After that the field was switched-off, samples were cooled down to room temperature without electric field, and the piezoelectric coefficient $d_{33}$ was measured Obtained results are analogous to results obtained on the samples with coexisting FE and AFE phases. The solid solutions become piezoelectrically active only if their content of Zr/Ti corresponds to the MPB region of the phase diagram. The is one more experiment indicative of the existence of domains of the AFE phase in solid solutions with compositions from the MPB region.

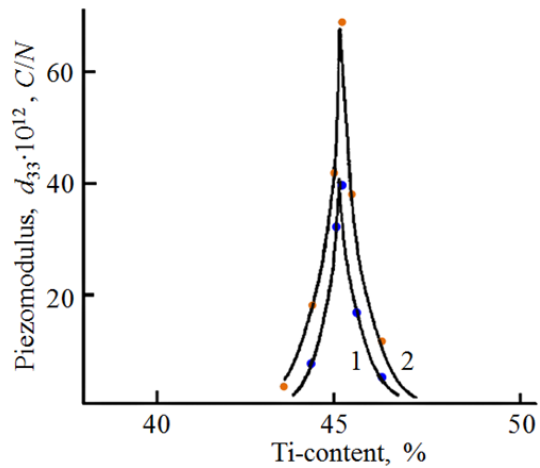

Fig.9. Piezoelectric modulus vs content of titanium in the $(Pb_{0.9}Ba_{0.05}Sr_{0.05})_{0.985}(Li_{0.5}La_{0.5})_{0.015}(Zr_{1-x}Ti_x)O_3$ solid solutions subjected to the thermoelectric treatment during time $\tau$ in paraelectric phase at temperature $T_C + 50$ °C in DC electric field of 120 V/mm.
Treatment time, $\tau$, min: 1 – 20; 2 – 40.

### D. Paraelectric phase transitions in an external electric field

Now let us dwell on another series of experiments that demonstrate the peculiarities in characteristic of substances with coexisting domains of the FE and AFE phases in the bulk of the samples. The point of the paraelectric phase transition in these materials shifts under the heating in the presence of an external electric field. The transition point in ferroelectrics is shifted toward the higher temperatures whereas in antiferroelectrics the transition point is shifted toward the lower temperatures. Such behavior is caused by a decrease in the energy of ferroelectrics and by an increase in the energy of antiferroelectrics in an applied electric field. The dipole ordering temperature (the transition point) does not depend on the field intensity in substances with coexisting domains of the FE and AFE phases. The redistribution of the shares of these phases (in favor of the FE phase) takes place in an applied field and the total energy of the system remains constant [14]. The so-called intermediate state is realized in an applied field when domains of the FE and AFE phases coexist in the sample's volume. An applied electric field shifts the FE-AFE interphase boundary in such a way that the share of the FE phase rises, but at the same time the internal state inside each domain remains unchanged. When the substance is in the intermediate state the energy of domains of each phase is independent on electric field and the total energy of the whole sample is independent on the field. Therefore, in this case the temperature of the phase transition into the paraelectric state is constant in a variable field.

The results of measurements of the dependence of the Curie point on the electric field for the $Pb_{0.90}(Li_{0.5}La_{0.5})_{0.10}(Zr_{1-x}Ti_x)O_3$ series of solid solutions located at different distance from the region separating the FE and AFE states in the phase diagram (given in Fig.4) are presented in Fig.10. As one can see, the temperature of the paraelectric phase transition does not depend on the electric field when the solid solution composition permits the realization of the intermediate state of coexisting FE and AFE domains in electric field.

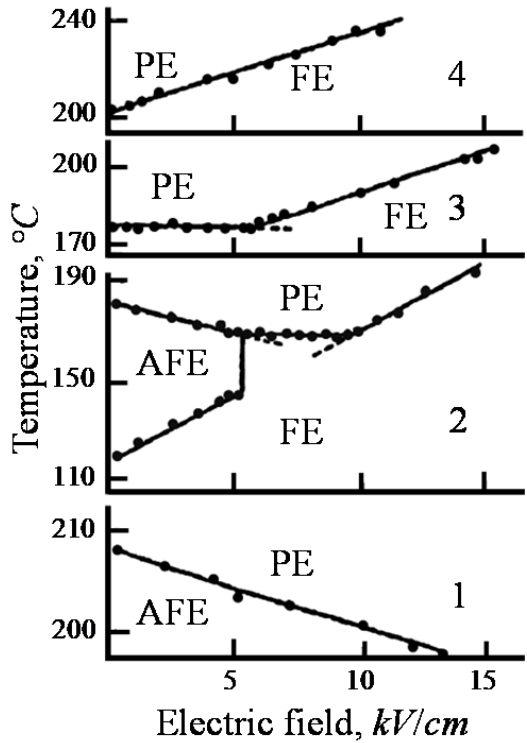

Fig.10. Dependencies of the value of Curie temperature on an external electric field intensity in the Pb$_{0.90}$(Li$_{0.5}$La$_{0.5}$)$_{0.10}$(Zr$_{1-x}$Ti$_x$)O$_3$ series of solid solutions from different regions of the phase diagram (phase diagram for this series of solid solutions is presented in Fig.4) characterized by different types of dipole ordering in low-temperature state.
The content of titanium, x, %: 1 – 10; 2 – 14; 3 – 18; 4 – 25.

Analogous results for the (Pb$_{0.9}$Ba$_{0.05}$Sr$_{0.05}$)$_{0.985}$(Li$_{0.5}$La$_{0.5}$)$_{0.015}$(Zr$_{1-x}$Ti$_x$)O$_3$ solid solutions are presented in Fig.11. In these solid solutions the peculiarities, typical for substances with the coexisting domains of the FE and AFE phase, are present only in the solid solutions belonging to the MPB region of compositions.

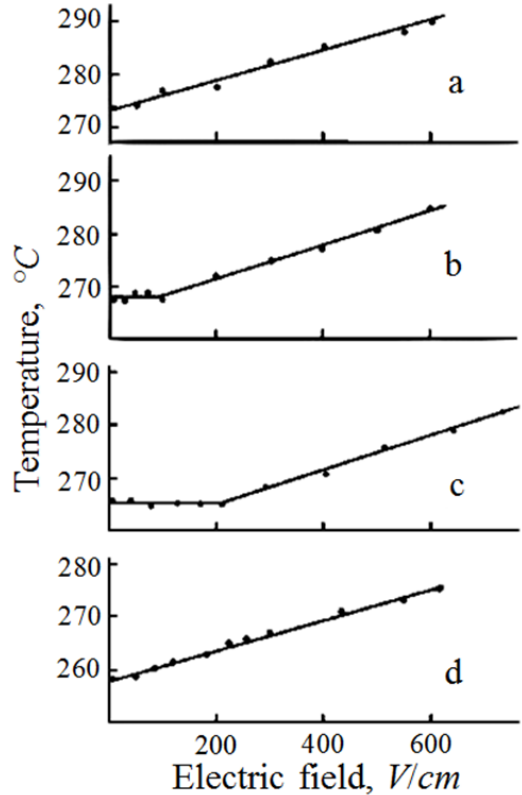

Fig.11. Dependencies of the value of Curie temperature on an external electric field intensity in the $(Pb_{0.9}Ba_{0.05}Sr_{0.05})_{0.985}(Li_{0.5}La_{0.5})_{0.015}(Zr_{1-x}Ti_x)O_3$ solid solutions.
The content of titanium, x, %: a – 47.0; b – 45.750; c – 45.50, d – 43.0.

This experiment also indicates the existence of domains of the AFE phase in the PZT-based solid solutions with compositions belonging to the MPB region (Fig.1).

Thus, we have demonstrated that the solid solutions with compositions from the MPB region have the same peculiarities in the behavior of physical characteristics as the solid solutions located in the vicinity of the FE-AFE phase boundary in the "composition-temperature" phase diagram. The MPB region is usually located far from this boundary and the peculiarities in behavior of such solid solutions are not considered to be connected with the possible presence of domains of the AFE states. Moreover the energies of such states are widely spaced, which allows not considering them in the solid solutions belonging to the MPB region. However, one's understanding of phenomena will be changed substantially if one takes into account the local decomposition of solid solutions in the vicinity of the tetragonal-rhombohedral interphase boundary.

In the frame of such approach one has to take into account that the inhomogeneity in distribution of zirconium and titanium ions in the vicinity of the interphase boundary can lead to appearance of nanodomains greatly enriched with zirconium. The AFE type of dipole ordering can exist inside such Zr-enriched nanodomains. The coexistence of such AFE domains with domains with the FE ordering defines the peculiarities in behavior of the investigated solid solutions. These nanodomains with the AFE ordering can be identified by the transmission electron microscopy.

## IV. DISCUSSION

Results presented in this work find unique and consistent explanation in the framework of concept about the coexistence of domains of the dipole ordered FE and AFE phases and physical processes occurring at the interphase boundaries between these domains. The AFE state alone is absolutely unfavorable in solid solutions belonging to the MPB region since the MPB region and the region of the AFE states are located quite far from one another in the "composition-temperature" phase diagram of phase states in PZT. The appearance of the AFE state is only possible as a consequence of additional factors, in particular, as a consequence of the local decomposition of solid solutions and zirconium enrichment of local regions in the vicinity of interphase boundaries (together with local mechanical stresses or without such stresses).

Let us consider data available in the literature that additionally argues for the point of view (based on experimental results of the section 3) of this article.

In the majority of contemporary considerations of transition from the rhombohedral phase to the tetragonal phase (in other words the analysis of the conjugation of phases) and the relaxation of possible stresses on the coherent boundaries between these phases when the domains of these phases coexist in the sample this transition is regarded as taking place by a rotation of the polarization vector. From the crystallographic point of view such rotation must lead to the appearance of the monoclinic phases [25, 27].

We suggest taking into consideration a new phase for explanation of structural peculiarities of substances belonging to the MPB regions of PZT-based solid solutions. Essentially, such approach is not new. Similar assumptions have been made earlier. An exception is only the circumstance that nobody suggested an appearance of the AFE phase domains in these solid solutions.

Articles [28, 29] were among the first that used transmission electron microscopy in the selected-area electron diffraction (SAED) mode. The SAED pattern observed in these papers showed the double and the triplet splitting (for high-index planes) of the diffraction pattern. Authors attributed the triplet spitting to the twin-related tetragonal (T), rhombohedral (Rh) adjacent FE domains. The interpretation of diffraction pattern was done on the base of succession of domains of the $T_1$-Rh-$T_2$-Rh-$T_1$ phases. That is the authors of [28, 29] consider that SAED points to a presence of one more phase with tetragonal distortions in solid solutions belonging to the MPB region of PZT.

Authors of [30-32] studied the PZT solid solutions belonging to MPB region by synchrotron radiation. They observed lines which in no way were related to the monoclinic phases. Using transmission electron microscopy they directly observed the nanodomains with sizes of the order of 10-15 nm. As judged by the photographs given in these papers the location of these nanodomains corresponds to the interphase boundaries. These boundaries are quite wide due to the changes in the distribution of ions in equivalent crystallographic positions in the vicinity of CIPB. Authors of [31] traced the behavior of these nanodomains in an applied electric field. The applied field transformed these nanodomains into the domains of FE rhombohedral phase. That is the field caused the induced phase transition into the FE state. Since the crystal structure of nanodomains is not cubic this transition is similar to the induced transition from the AFE state into the FE state.

The results of Ref. 33 also strongly suggest possible presence of the AFE states in the solid solutions from the MPB region. The crystal structure of micron-sized domains and nanodomains was investigated by transmission electron microscopy. Authors of ref 33 attributed the complex domain structures observed at $x = 0.5$ and $0.95$ (the last solid solution is located inside the FE-AFE morphotropic region) to a physical manifestation of the frustration between and Zr- and Ti-rich clusters.

Results pointing to possible existence of the AFE phase domains in PZT-based solid solutions with compositions belonging to the MPB region have been obtained in [34] by inelastic X-ray scattering. The measurement revealed that upon cooling from the paraelectric phase, the spectral response of the soft $M_{5'}$ zone boundary phonon mode associated with antiferroelectric vibrations of lead ions existed and is progressively transformed to a broad central mode. The authors believe that the behavior observed in their experiments directly related to the nanoscale inhomogeneity of the structure.

It is necessary to draw ones attention to one more peculiarity of properties of the PZT-based solid solutions belonging to the MPB region. Neutron diffraction, spectroscopic, as well as electron diffraction experiments indicated the presence of displacements of ions that correspond to the rotational modes of the oxygen octahedral with wave vector at the Brillouin zone boundary. As a rule researches do not relate these modes and ion displacements to the presence of the AFE phase in these solid solutions. The authors of [35] carried out density functional calculations in the Hedin-Lundqvist local density approximation using a general potential linearized augmented plane-wave method, with local orbital extensions to relax linearization errors and treat semi-core states to investigate the competitions between ferroelectric and rotational instabilities in rhombohedral PZT from the MPB region. They found a ferroelectric instability and also found a substantial $R$-point rotational instability, close to the ferroelectric one. The existence of both these instabilities close to one another is similar to the situation in pure $PbZrO_3$. As it was also noted in Ref. 30 these two instabilities are both strongly pressure dependent, but in opposite directions so that lattice compression of less than 1% is sufficient to change their ordering. Authors of Ref. 35 came to the conclusion that due to this, and local stress fields due to the $B$-site cation disorder may lead to coexistence of both types of instability are likely present in the MPB region. It is again quite similar to the situation in pure $PbZrO_3$.

Results of our studies given in section 3.3 (and also in section 3.2) obtained at the temperatures above the Curie temperature of solid solutions find their explanation if the complex (FE+AFE) domains exist in the paraelectric phase in the vicinity of the $Rh_{FE} - T_{FE} - PE_C$ triple point in the phase diagram. In our opinion the results of [34] support our approach. Dispersion curves of the lowest frequency transverse optic and the transverse acoustic phonon modes propagating along the [100] direction have been determined in the high-temperature paraelectric phase as well as at the room-temperature. Recall that the spectral response of the zone boundary phonon modes obtained in this paper is associated with the antiferroelectric vibrations of lead ions.

As we have already stated in the beginning of this article and want to stress one more time that we are not trying to prove that the monoclinic phases do not exist in the PZT-based solid solutions with compositions belonging to the MPB region of the phase diagram. We are simply trying to draw ones attention to the fact that not all the factors were taken into account during the interpretation of experimental results. It is primarily relevant to the results obtained by the treatment of the profiles of the lines in the scattering spectra of different nature – X-ray, neutron, and optic.

Let us give one more argument for our point of view. A large number of publications substantiating the possibility of development of piezoelectric materials with large values of piezoelectric parameters appeared after the publication of first papers about monoclinic phases. And what kind of result achieved in this direction during last 15-16 years? PZT-based ceramic materials had the largest value of piezoelectric coefficient $d_{33}$ at the level of $560 \cdot 10^{-12}$ - $600 \cdot 10^{-12}$ $pC/N$ before the time when the idea about monoclinic phase appeared. Using the monoclinic phase idea one could improve the value of $d_{33}$ up to ~ $630 \cdot 10^{-12}$ pC/N. However, in the framework of idea about the local decomposition of the solid solutions near the CIPB between the rhombohedral and tetragonal phases and the idea about the formation of nanodomains of the AFE phase the piezoelectric coefficient value of more than $1100 \cdot 10^{-12}$ pC/N (see Fig. 8) has been achieved. As one can see the results are disparate. But this is not the most interesting circumstance. We have obtained the value of $d_{33}$ piezoelectric modulus equal to $1200 \cdot 10^{-12}$ pC/N in the hard ferroelectric PZT-based ceramics with composition from the MPB region and a coercive field of 1900 V/mm. These results demonstrate one more time that not all the physical effects have been taken into account while interpreting experiments.